\documentclass[prb,twocolumn,superscriptaddress,a4paper,twoside,showpacs]{revtex4}

\usepackage{amsmath}
\usepackage{amssymb}
\usepackage{todonotes}
\usepackage{nicefrac}
\usepackage{graphics}
\usepackage{color}

\makeatletter
\newcommand*{\rom}
[1]{\expandafter\@slowromancap\romannumeral #1@}
\makeatother
\renewcommand{\vec}[1]{\ensuremath{\mathbf{#1}}}
\renewcommand{\Im}{\ensuremath{\operatorname{Im}}}


\begin{document}

\title{When strong correlations become weak: Consistent merging of $GW$ and DMFT}
\author{L. Boehnke}
\email{lewin.boehnke@unifr.ch}
\affiliation{Department of Physics, University of Fribourg, 1700 Fribourg, Switzerland}
\author{F. Nilsson}
\email{fredrik.nilsson@teorfys.lu.se}
\affiliation{Department of Physics, Division of Mathematical Physics, Lund University, S{\"o}lvegatan 14A, 223 62 Lund, Sweden}
\author{F. Aryasetiawan}
\affiliation{Department of Physics, Division of Mathematical Physics, Lund University, S{\"o}lvegatan 14A, 223 62 Lund, Sweden}
\author{P. Werner}
\affiliation{Department of Physics, University of Fribourg, 1700 Fribourg, Switzerland}

\begin{abstract}
The cubic perovskite SrVO$_3$ is generally considered to be a prototype strongly correlated metal with a characteristic three-peak structure of the $d$-electron spectral function, featuring a renormalized quasiparticle band in between pronounced Hubbard sidebands.
Here we show that this interpretation, which has been supported by numerous ``ab-initio'' simulations, has to be reconsidered.
Using a fully self-consistent $GW$+extended dynamical mean-field theory calculation we find that the screening from nonlocal Coulomb interactions substantially reduces the effective local Coulomb repulsion, and at the same time leads to strong plasmonic effects.
The resulting effective local interactions are too weak to produce pronounced Hubbard bands in the local spectral function, while prominent plasmon satellites appear at energies which agree with those of the experimentally observed sidebands.
Our results demonstrate the important role of nonlocal interactions and dynamical screening in determining the effective interaction strength of correlated compounds.
\end{abstract}

\maketitle

SrVO$_{3}$ has been considered a prototype strongly correlated metal ever since photoemission and inverse photoemission experiments twenty years ago~\cite{Morikawa95Spectral} showed features at energies well outside the renormalized  quasiparticle band.
These satellites have been explained as Hubbard bands, because they appear in combined density functional + dynamical mean-field theory (LDA+DMFT)~\cite{Georges96Dynamical} simulations when the local Coulomb repulsion is chosen such that the experimentally observed mass renormalization is reproduced (see e.g. Refs.~\onlinecite{Pavarini04Mott,Lechermann06Dynamical,Backes16Hubbard}).
Comparable values for the ``Hubbard $U$'' on the order of $5$ eV were obtained by constrained LDA~\cite{Nekrasov06Momentumresolved, Taranto13Comparing} and used in DMFT calculations with static local interactions.
The constrained random phase approximation (cRPA)~\cite{Aryasetiawan04Frequencydependent} provides a systematic way of computing the dynamically screened interaction parameters consistent with the LDA bandstructure, and the resulting local $U(\omega)$ of the DMFT auxiliary system can be efficiently handled by state-of-the-art impurity solvers~\cite{Werner10Dynamical}.
These more realistic calculations however produce a too strong renormalization of the quasiparticle band~\cite{Sakuma13Electronic}.
The missing ingredients in the LDA+DMFT+$U(\omega )$ approach are the nonlocal selfenergy and polarization effects, and the additional screening of the $U(\omega )$ resulting from nonlocal Coulomb interactions within the low-energy subspace.

A promising scheme, which can treat all these effects in a consistent manner, is the combination of the $GW$ ab-initio method~\cite{Hedin65New} and extended DMFT (EDMFT)~\cite{Sun02Extended,Biermann03FirstPrinciples}.
While this $GW$+EDMFT formalism has been tested on simple one-band Hubbard models~\cite{Karlsson05Selfconsistent,Ayral13Screening,Huang14Extended,Hansmann13LongRange}, and several simplified versions have been applied to SrVO$_{3}$~\cite{Tomczak12Combined,Sakuma13Electronic,Taranto13Comparing,Tomczak14Asymmetry}, a fully self-consistent implementation in an ab-initio setting has so far been hampered by the challenges of solving the bosonic self-consistency loop for multiorbital systems and nontrivial issues related to a proper embedding of the EDMFT calculations into a $GW$ ab-initio framework.

\begin{figure}[b]
\includegraphics{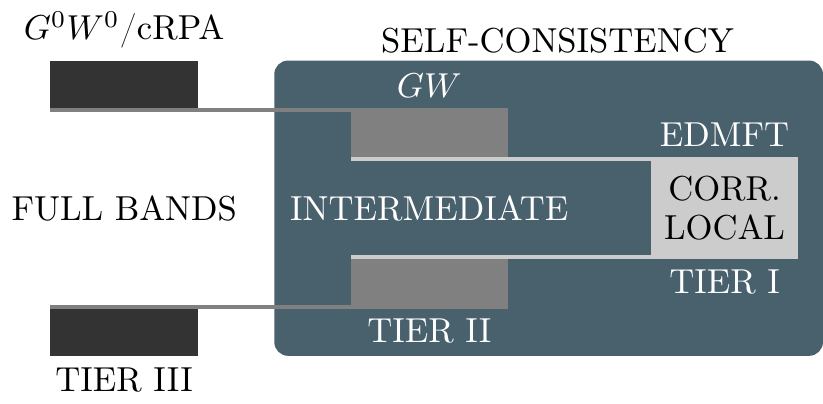}

\caption{Different levels of approximation in the multi-tier $GW$+EDMFT scheme: The LDA+$G^0W^0$ treatment of the full range of bands (tier~\rom{3}) is used to construct the effective model for the intermediate energy space of Wannier functions (tier~\rom{2}), which is handled in $GW$.
An effective model for the local part (potentially further reduced in number of orbitals) of tier~\rom{2} is constructed and handled as the EDMFT impurity problem (tier~\rom{1}).
Tiers~\rom{1} and~\rom{2} are solved self-consistently.}
\label{fig:GWEDMFTtiers}
\end{figure}

For a consistent description of the physics of a realistic material we have to employ a multi-tier approach that handles orbitals and interactions with appropriate approximations (Fig.~\ref{fig:GWEDMFTtiers}).
Starting from a density functional theory (DFT) calculation in the local density approximation (LDA)~\cite{Kohn65SelfConsistent} for the full range of bands in the solid, we first perform a one-shot $GW$ calculation ($G^{0}W^{0}$).
We then define an intermediate subspace $I$, for which the goal is to construct an accurate low-energy model.
This requires a suitable $I$~subspace approximation for the bare fermionic and bosonic propagators.
Furthermore, we introduce a possibly smaller correlated orbital subspace $C$, whose effective local interactions are treated by means of an impurity construction within an extended DMFT approach similar to Refs.~\onlinecite{Ayral13Screening,Huang14Extended,Tomczak12Combined}.
The fermionic and bosonic self-consistency loops are solved with bare propagators that incorporate the effect of screening channels outside $I$, while the self-consistently determined hybridization functions and retarded interactions of the impurity model also include retardation effects originating from nonlocal interactions and screening processes within $I$.
For SrVO$_3$ with one electron in the $d$ orbitals, we choose the $t_{2g}$ subspace for both $I$ and $C$, because of the clear-cut energy separation.

For the electron gas, it is known that fully self-consistent $GW$ calculations produce accurate total energies but the quasiparticle dispersion has been found to be unsatisfactory: the self-consistent occupied band width is wider than the one-shot result, worsening agreement with experiment \cite{Holm98Fully}.
In the present self-consistent scheme, we restrict self-consistency to the intermediate subspace in which vertex corrections beyond the $GW$ approximation are supplied by the EDMFT.
As we will show, these vertex corrections counteract the undesirable effects of self-consistency within a pure $GW$ approximation.

\begin{figure}
\begin{center}
\includegraphics{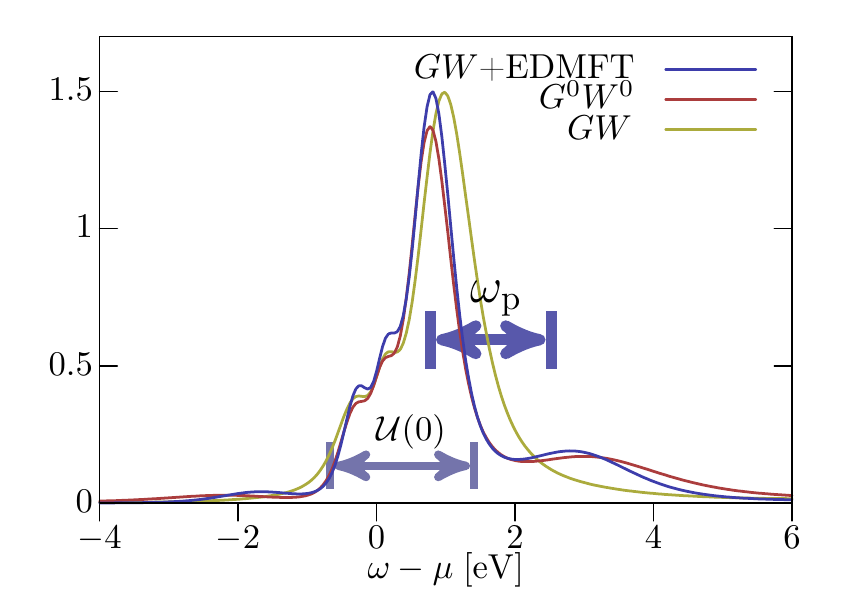}
\end{center}
\caption{Local spectral function of SrVO$_3$ calculated in different approximations:
Single shot $G^0W^0$, self-consistent $GW$ and $GW$+EDMFT.}
\label{fig:SrVO3doses}
\end{figure}

In the multi-tier approach, the total Green's function evaluated in $I$ can be expressed as 
\begin{align}
G_{\vec{k}}^{-1}=&\overbrace{\underbrace{\mathrm{i}\omega _{n}+\mu -\varepsilon _{\vec{k}}^{\mathrm{LDA}}+V_{\mathrm{XC},\vec{k}}}_{G_{\mathrm{LDA},\vec{k}}^{0}{}^{-1}}\underbrace{-\Sigma_{\vec{k}}^{G^{0}W^{0}}+\Sigma_{\vec{k}}^{G^{0}W^{0}}\big|_{I}}_{-\Sigma _{\mathrm{r},\vec{k}}\big|_{I}}}^{\mathrm{TIER~\rom{3}},\; G_{I,\vec{k}}^{0}{}^{-1}}\notag\\
&\underbrace{-\Sigma _{\vec{k}}^{GW}\big|_{I}+\Sigma ^{GW}\big|_{C,\mathrm{loc}}}_{\mathrm{TIER~\rom{2}}}\underbrace{-\Sigma ^{\mathrm{EDMFT}}\big|_{C,\mathrm{loc}}}_{\mathrm{TIER~\rom{1}}}\;,\label{eqn:fullG}
\end{align}
where $G_{\mathrm{LDA},\vec{k}}^{0}$ is the LDA bare propagator and $\Sigma_{\mathrm{r},\vec{k}}\big|_{I}$ the effect of the rest  space (everything except for $I$) on the intermediate Green's function.
A Matsubara-frequency dependence is implicit for all $G$ and $\Sigma$ objects.
The notation $A\big|_{B}$ implies that also all internal processes in the evaluation of $A$ are restricted to the space $B$.
$V_{\mathrm{XC}}$ is the exchange-correlation potential of the LDA calculation.
The object marked $G_{I,\vec{k}}^{0}$ plays the role of an effective bare propagator in a Dyson equation for the effective $I$~model, which is solved self-consistently.
$G^0_{I,\vec{k}}$ excludes contributions from the one-shot $G^0W^0$ to the intermediate subspace.
This is replaced by the second line in Eq.~\eqref{eqn:fullG}, which then plays the role of the selfenergy in the Dyson equation with $G^0_{I,\vec{k}}$ as the bare propagator.
The selfenergy in that Dyson equation is separated into the $GW$ selfenergy $\Sigma_{\vec{k}}^{GW}\big|_{I}$, and the extended DMFT impurity selfenergy $\Sigma^{\mathrm{EDMFT}}\big|_{C,\mathrm{loc}}$.
The part of the selfenergy that is included in both, $\Sigma^{GW}\big|_{C,\mathrm{loc}}$, is what would conventionally be called the double counting $\Sigma_{\mathrm{DC}}$.
In contrast to the LDA+DMFT scheme, this term is unambiguously defined within the $GW$+EDMFT framework.

The analogous expressions for the bosonic propagators are
\begin{align}
W_\vec{q}^{-1}=&\overbrace{v_\vec{q}^{-1}\underbrace{-\Pi^{G^0G^0}_\vec{q}+\Pi^{G^0G^0}_\vec{q}\big|_I}_{-\Pi_{\mathrm{r},\vec{q}}}}^{\mathrm{TIER~\rom{3}},\; U_{I,\vec{q}}^{-1}}\notag\\
&\underbrace{-\Pi_\vec{q}^{GG}\big|_I+\Pi^{GG}\big|_{C,\mathrm{loc}}}_{\mathrm{TIER~\rom{2}}}\underbrace{-\Pi^\mathrm{EDMFT}\big|_{C,\mathrm{loc}}}_{\mathrm{TIER~\rom{1}}}\;,\label{eqn:fullW}
\end{align}
where $v_\vec{q}$ is the bare Coulomb interaction and $\Pi_{\mathrm{r},\vec{q}}$ the screening effect on the $I$~interaction at the cRPA~\cite{Aryasetiawan04Frequencydependent} level.
Consequently, $U_{I,\vec{q}}$ plays the role of a bare interaction for the $I$ subspace, in a bosonic Dyson equation in which the polarization $\Pi$ is built from $\Pi_q^{GG}\big|_I$ on tier~\rom{2} and $\Pi^\mathrm{EDMFT}\big|_{C,\mathrm{loc}}$ on tier~\rom{1}, again removing the part $\Pi^{GG}\big|_{C,\mathrm{loc}}$ that is included in both tiers.

Our approach is a consistent ab-initio $GW$+EDMFT implementation because it combines (i) a realistic starting point for both, the hopping processes and the local and nonlocal interactions, (ii) a sound treatment of the slow $\nicefrac{1}{r} $ decay of the Coulomb interaction,  (iii) the retardation effects resulting from the high-energy subspace, and (iv) the handling of the frequency-dependence of the interaction in the $GW$ calculation and the impurity problem.
Apart from the definition of the different subspaces, the simulation does not involve any free parameters.
While previous studies have properly treated some of the above-mentioned points, this work presents the first fully consistent ab-initio simulation of a material within $GW$+EDMFT.
It is also noteworthy that, since we perform the calculations self-consistently, the double-counting term is unambiguously defined.
This is not the case in a non-self-consistent scheme, in which the double-counting term depends on whether it is removed from the $GW$ or EDMFT contribution.

It would be straightforward in principle to extend the self-consistency to the full set of bands in tier~\rom{3} by updating the LDA density with the correlated density from the self-consistent tier~\rom{1}+tier~\rom{2} calculation in an outer self-consistency loop, similar to the charge self-consistent treatment developed for LDA+DMFT calculations~\cite{Pourovskii07Selfconsistent,Amadon08Planewave,Grieger12Approaching,Leonov15Metalinsulator}, but we believe that the current level of self-consistency is sufficient provided that $I$ is chosen large enough that correlations in the local correlated subspace do not lead to changes of electronic properties outside this subspace.

\begin{figure}
\begin{center}
\includegraphics{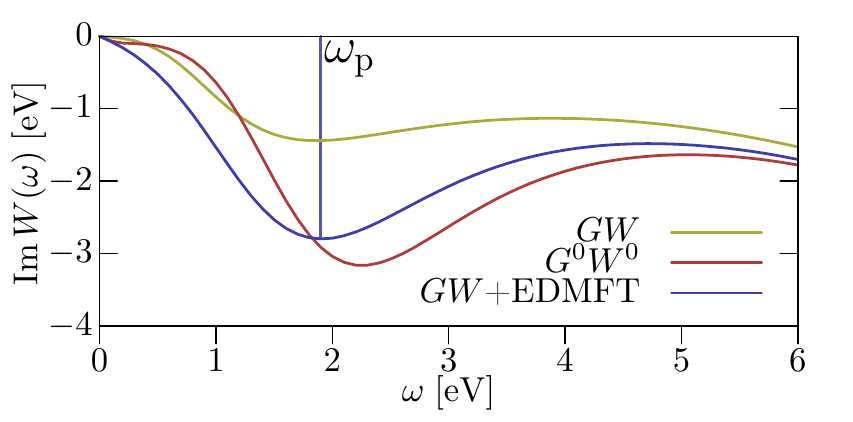}
\end{center}
\caption{Fully screened interaction for SrVO$_3$ in one-shot $G^0W^0$, self-consistent $GW$ and $GW$+EDMFT.}
\label{fig:SrVO3Ws}
\end{figure}

The LDA+$G^0W^0$ calculations on tier~\rom{3} were performed using the full-potential linearized augmented planewave tools \textsc{Fleur}/\textsc{Spex}~\cite{Fleur,Friedrich10Efficient} with 8x8x8 k-points and 200 bands for both the polarization function and the selfenergy.
From the LDA bandstructure the $I/C$-subspaces (both V $t_{2g}$ in this case) were defined using maximally localized Wannier functions (MLWF's) as implemented in the \textsc{WANNIER90} library~\cite{Marzari97Maximally,Souza01Maximally,Mostofi08Wannier90,Freimuth08Maximally,Sakuma13Symmetryadapted}.
MLWF's provide a convenient basis for the subsequent self-consistency cycle in tiers~\rom{1} and~\rom{2}.
In tier~\rom{2} the $GW$-selfenergy $\Sigma_\vec{k}^{GW}\big|_I$ and the corresponding double-counting term $\Sigma^{GW}_\vec{k}\big|_{C,\mathrm{loc}}$ were computed using a custom finite-temperature self-consistent $GW$ implementation including local vertex corrections to the screened interaction $W$ (Eq.~(\ref{eqn:fullW})).
Tier~\rom{1} was solved using the \textsc{Alps}~\cite{Bauer11Alps,Alps} implementation~\cite{Hafermann13Efficient} of the CT-Hyb algorithm~\cite{Werner06ContinuousTime,Gull11Continuoustime,Boehnke11Orthogonal}, which can handle the retarded impurity interactions~\cite{Werner10Dynamical} for the density-density components of the interaction tensor.
The self-consistency cycle was implemented on the $t_{2g}$ manifold for the Green's function and a full product basis thereof for the fully screened interaction using the \textsc{Triqs} framework~\cite{Parcollet15Triqs}.
All calculations were performed at $\beta=15\frac{1}{\mathrm{eV}}$.
The analytic continuation of the Matsubara frequency data was done using MaxEnt~\cite{Bryan90Maximum,Huang14Extended}, with an additional temperature broadening applied to the fermionic spectra. More details can be found in~\onlinecite{supp}.

In Fig.~\ref{fig:SrVO3doses} we show how the $t_{2g}$ spectral function obtained from our self-consistent $GW$+EDMFT simulations  compares to single shot $G^0W^0$ and self-consistent $GW$ calculations.
The $G^0W^0$ result has been discussed before~\cite{Sakuma13Electronic,footnote_sakuma}.
It exhibits a three-peak structure, but with satellite energies that are too high compared to experiment.
Self-consistent $GW$ substantially worsens the agreement with experiment~\cite{Holm98Fully,Aryasetiawan98GW}, leading to a broadening of the quasiparticle band and a washing-out of the satellites.
The $GW$+EDMFT spectrum exhibits the signature three-peak structure reminiscent of previous LDA+DMFT results, with correct satellite positions.
The physical picture which emerges is however new:  
{\it SrVO$_3$ turns out to be a weakly correlated metal, with rather low static local interactions, but with pronounced plasmonic satellites due to screening processes within the low-energy space $I$.}

The reduced static interaction (compared to the static cRPA value of $3.4$ eV) and the formation of satellite features in the spectral functions are a consequence of nonlocal interactions, which are reflected in the extended DMFT impurity problem by the self-consistently determined retarded on-site interaction $\mathcal{U}(\omega)$ \cite{Ayral13Screening}.
Marked in Fig.~\ref{fig:SrVO3doses} is the static intraorbital value of this interaction, $\mathcal{U}(\omega=0)=2.2$ eV, which is clearly too weak to account for the $4.5$ eV separation between the side-peaks in a Mott picture.
Instead, these sidebands can be explained as plasmon \cite{footnote_plasmon} satellites of the quasiparticle band, generated by a retardation channel in the fully screened interaction at $\omega _{\mathrm{p}}$, as seen in Fig.~\ref{fig:SrVO3Ws}.
These observations are not sensitive to the details of the quasiparticle structure~\cite{supp}.
Compared to the $G^0W^0$ result, the $GW$+EDMFT quasiparticle structure is hardly modified, while the satellites are shifted closer to the Fermi energy.

\begin{figure}[t]
\begin{center}
\includegraphics{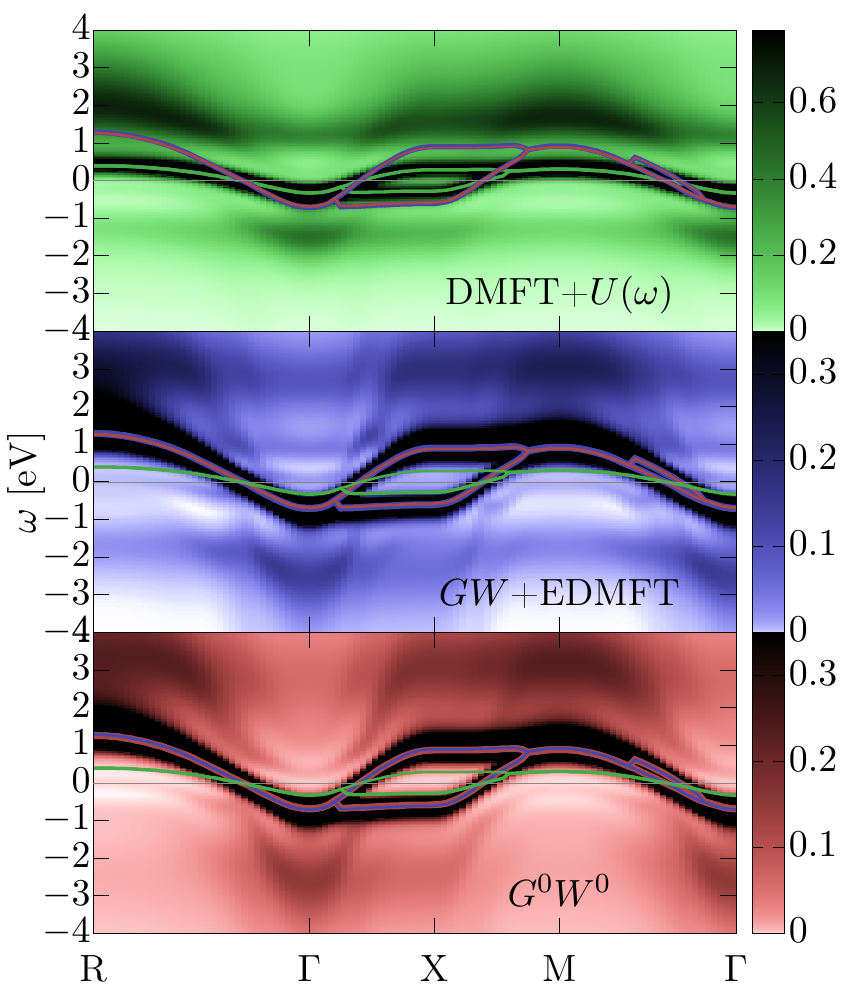}
\end{center}
\caption{$\vec{k}$-resolved spectral function from DMFT with screened local $U(\omega)$ taken from cRPA, $GW$+EDMFT and one-shot $G^0W^0$.
For easier comparison, lines mark the maxima of the spectra.}
\label{fig:SrVO3Akw}
\end{figure}

As shown by the $\vec{k}$-resolved spectral functions in Fig.~\ref{fig:SrVO3Akw}, the dispersion of the $GW$+EDMFT satellites is clearly resembling the $G^0W^0$ result, with additional rather flat structures in the $GW$+EDMFT spectra that might have Hubbard band character.
If we identify the flat structure in the occupied part with the lower Hubbard band, then the upper Hubbard band has to be located near the upper edge of the quasiparticle band, 
which means that it cannot be identified in momentum-integrated spectra. 
In the local spectral function, the satellite features are pulled to $-1.7$ eV and $2.8$ eV, in good agreement with experimental data~\cite{Morikawa95Spectral,Sekiyama04Mutual,Yoshida10Mass}.
The fact that the lower satellite is rather weak and dominated by the plasmon contribution is consistent with the results of Ref.~\onlinecite{Backes16Hubbard}, which showed that oxygen vacancies also give rise to spectral weight around $-1.5$ eV.
It is thus likely that the weight of the lower satellite was overestimated in previous experiments.
The peculiar intensity modulation with increased weight around the $\Gamma$ point was also experimentally observed before~\cite{Takizawa09Coherent}.

In Ref.~\onlinecite{Gatti13Dynamical} the spectral function for SrVO$_3$ was calculated within the quasiparticle self-consistent $GW$ approximation \cite{Schilfgaarde06Quasiparticle,Bruneval06Effect} with the cumulant expansion for the Green's function (QPSC$GW$+C).
A similar line of reasoning was previously presented in Ref.~\onlinecite{Gatti07Understanding} in the context of VO$_2$, where the satellites are also naturally identified as plasmons.
The QPSC$GW$+C spectra agree remarkably well with the present calculation, especially in the unoccupied part of the spectra. 
The main difference between the $GW$+EDMFT spectrum in Fig.~\ref{fig:SrVO3doses} and the QPSC$GW$+C spectrum in Ref.~\onlinecite{Gatti13Dynamical} is the position of the satellite in the occupied part of the spectrum which is virtually the same as in plain $G^0W^0$ in the latter case.
Hence, SrVO$_3$ is a delicate material with strong nonlocal screening effects so that, on the one hand, a purely local treatment within LDA+DMFT overestimates the correlations and provides a wrong physical interpretation of the spectra while, on the other hand, local vertex corrections beyond $GW$ are essential to reproduce the photoemission data.

The remarkable agreement between the cumulant expansion and the present result gives further support of our interpretation of SrVO$_3$ as a weakly correlated material.
The cumulant expansion is an effective method for improving high-energy satellite features by including diagrams with multiple emissions and absorptions of plasmons whereas quasiparticle features are relatively untouched, as reflected in the negligible improvement over the $G^0W^0$ result on the quasiparticle dispersion.
The cumulant expansion, however, is based on a many-body perturbation expansion and does not capture the strong local correlations responsible for shifting the lower satellite closer to the Fermi level.

The effect of the local vertex corrections to the selfenergy and polarization become clear from the comparison of the different approximations in Fig.~\ref{fig:SrVO3doses}.
Without the vertex corrections from EDMFT the current scheme reduces to a self-consistent $GW$ calculation within the $I$-subspace.
In this case, the quasiparticle bandwidth is broadened and the satellite features are washed out, similar to what has been found previously for the electron gas \cite{Holm98Fully}.
Looking at Fig.~\ref{fig:SrVO3Ws}, it is interesting to note that the vertex corrections restore the weight and width of the self-consistent $GW $ plasmon peak to the one-shot $G^0W^0$ result whereas the energy is only slightly increased.

The reduction of the plasma frequency in self-consistent $GW$ compared to the one-shot result can be understood from the reduction of low-energy spectral weight in the self-consistent Green's function.
While $W^0$ is calculated from the LDA bandstructure, the self-consistent $W$ is calculated from the interacting Green's function where part of the low energy spectral weight has been shifted to higher frequencies.
The increase in weight of the spectral function at high energy around the plasmon energy results in the broadening of the plasmon width and the reduction in its weight and energy, as also found in the case of the electron gas \cite{Holm98Fully}.
In the full $GW$+EDMFT calculations the non-trivial interplay between local vertex corrections and spectral weight reduction restores the weight of the pole in $\Im W(\omega)$ and at the same time shifts it to slightly higher energy.
This result clearly demonstrates that local vertex corrections are indeed essential to counteract the undesirable effects in self-consistent $GW$.

To conclude, we have demonstrated that the characteristic three-peak structure in SrVO$_3$ that was previously attributed to Mott physics is also found in more sophisticated self-consistent and parameter-free $GW$+EDMFT calculations.
It originates however from a different physical mechanism.
The local density of states has satellites at a separation that cannot be reconciled with the effective local interaction, since the latter is significantly reduced from $U_\text{cRPA}(\omega=0)=3.4$ eV to $\mathcal{U}(\omega=0)=2.2$ eV by screening through nonlocal processes in the EDMFT self-consistency loop.
The low-frequency quasiparticle structure in our self-consistent $GW$+EDMFT approach has a much closer resemblance to the single-shot $G^0W^0$ result than the self-consistent $GW$ calculation, an interesting observation in connection with the empirical fact that $G^0W^0$ often captures experimental findings very well~\cite{Aryasetiawan98GW}.
While $GW$+EDMFT lacks nonlocal self-energy contributions beyond $GW$, we expect that the new physical picture introduced here applies to other correlated metals in which the electron filling and long-range Coulomb interaction result in substantial nonlocal screening.

\begin{acknowledgments}
L.B. and P.W. acknowledge financial support from SNSF through NCCR MARVEL.
F.N. and F.A. acknowledge financial support from the Swedish Research Council (VR).
The computations were performed on resources provided by the Swedish National Infrastructure for Computing (SNIC) at LUNARC and at the CSCS Dora cluster provided by MARVEL.
L.B. and P.W. thank Denis Gole\v{z}, Nicola Marzari and David O'Regan for fruitful discussions.

L.B. and F.N. contributed equally to this work.
\end{acknowledgments}


\begin{thebibliography}{54}
\bibitem[{\citenamefont{{Note that in there the one-shot $G^0W^0$ was denoted
  by $GW$.}}()}]{footnote_sakuma}
\bibinfo{author}{\bibnamefont{{Note that in there the one-shot $G^0W^0$ was
  denoted by $GW$.}}}
\end{thebibliography}
\end{document}


\title{When strong correlations become weak: Consistent merging of $GW$ and DMFT\\Supplemental Material}
\author{L. Boehnke}
\email{lewin.boehnke@unifr.ch}
\affiliation{Department of Physics, University of Fribourg, 1700 Fribourg, Switzerland}
\author{F. Nilsson}
\email{fredrik.nilsson@teorfys.lu.se}
\affiliation{Department of Physics, Division of Mathematical Physics, Lund University, S{\"o}lvegatan 14A, 223 62 Lund, Sweden}
\author{F. Aryasetiawan}
\affiliation{Department of Physics, Division of Mathematical Physics, Lund University, S{\"o}lvegatan 14A, 223 62 Lund, Sweden}
\author{P. Werner}
\affiliation{Department of Physics, University of Fribourg, 1700 Fribourg, Switzerland}

\maketitle

\subsection{Nature of the satellites}
\begin{figure}
\begin{center}
\includegraphics{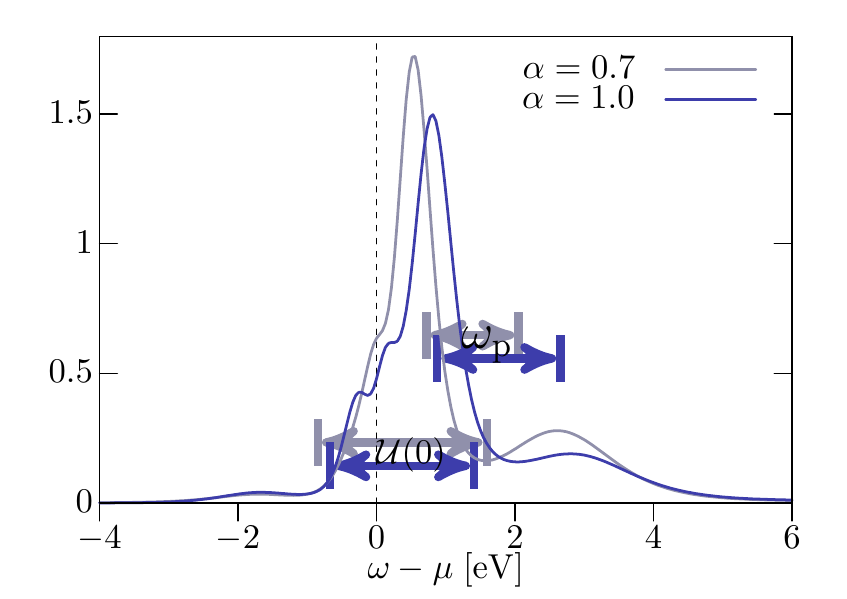}
\end{center}
\caption{Local spectral function of SrVO$_3$ in $GW$+EDMFT with ($\alpha=0.7$) and without ($\alpha=1.0$) scaling of the free propagator.}
\label{fig:SrVO3scaleddoses}
\end{figure}

To get additional information on the nature of the satellites and to demonstrate the robustness of our physical interpretation of the upper satellite, we compare the result in the main text to calculations in which the bandwidth is artificially reduced.
For this purpose, we introduce a scaling parameter $\alpha$ into the definition of the bare LDA propagator used in Eq.~(1),
\begin{equation}
  G^0_{\mathrm{LDA},\vec{k}}{}^{-1}=\operatorname{i}\omega_n+\mu-\alpha\varepsilon_\vec{k}^\mathrm{LDA}+V_{\mathrm{XC},\vec{k}}\quad.
\end{equation}
With this definition $\Sigma _{\vec{k}}^{GW}\big|_{I}$ and $\Pi_\vec{q}^{GG}\big|_I$ in Eq.~(1) are evaluated with the rescaled $t_{2g}$-bandwidth while $\Sigma _{\mathrm{r},\vec{k}}\big|_{I}$ and $\Pi_{\mathrm{r},\vec{q}}$ are evaluated with the unscaled bandwidth, since the two latter quantities are assumed to be insensitive to changes within the $I$-subspace.

The scaling parameter allows to phenomenologically include interaction effects and screening channels that are ignored in the current approach, since it will influence both the bare propagator on the $t_{2g}$ subspace and in turn the nonlocal screening (a completely flat band will only include local screening processes).
In particular, we note that the $G^0W^0$ approximation that is used outside $I$, and the restriction of $C$ to the $t_{2g}$ subspace (it would be preferable to use a better approximation also for the V $e_g$ states) affects the effective bare propagator on the $t_{2g}$-subspace.
A further approximation is the restriction of vertex corrections to local ones within $C$ which mainly affects the nonlocal screening.

Figure~\ref{fig:SrVO3scaleddoses} shows the local spectral function for $\alpha=1.0$ (the result from Fig. 2 in the main manuscript) and for $\alpha=0.7$, which yields a  quasiparticle bandwidth close to the experimentally observed one~\cite{Backes16Hubbard}. 
Reducing the bandwidth also reduces the plasmon frequency $\omega_\mathrm{p}$, while the effective static impurity interaction $\mathcal{U}(\omega=0)$ increases.
We find that the upper satellite shifts to lower energies, i.e., it follows the trend of $\omega_\mathrm{p}$ and opposes the trend in $\mathcal{U}(\omega=0)$, in agreement with our physical interpretation as a plasmonic feature, rather than an upper Hubbard band.

\subsection{Analytic continuation}
To obtain the electronic (Figs.~2,~4,~\ref{fig:SrVO3scaleddoses}) and bosonic (Fig.~3) spectral functions we use the maximum entropy (MaxEnt) method~\cite{Bryan90Maximum,Gubernatis91Quantum} and a code which is available online~\cite{MaxEnt}.
We checked the result against Pad\'e analytic continuation~\cite{Vidberg77Solving}, finding qualitative agreement, but chose the MaxEnt spectra for two main reasons.
Firstly, using MaxEnt analytic continuation has --in the presence of stochastic noise in the data-- a significantly better protection against deceptive spectral features due to its construction as the least biased spectrum in the absence of clear evidence.
Secondly, Pad\'e tends to represent well the position and weight of low-energy features in the spectral functions, but can be unstable with respect to their width.
In the present case, we are interested in resolving satellites that lie on the shoulders of the main quasiparticle structure, so that a change in the width results in a change in the position of the maximum, which complicates the physical interpretation.

For the analytic continuation of the bosonic spectra, we employ the method outlined in Ref.~\onlinecite{Huang14Extended}.
We rewrite
\begin{align}
  W(\tau)=&\int_{-\infty}^\infty\operatorname{d}\omega\frac{\exp{-\tau\omega}}{1-\exp{-\beta\omega}}\left(\frac{-\Im W(\omega)}{\pi}\right)\\
=&\int_0^\infty\operatorname{d}\omega K(\omega,\tau)\widetilde W(\omega)\label{eqn:bosonmaxent}
\end{align}
in terms of a kernel
\begin{equation}
K(\omega,\tau)\!=\!\frac{\operatorname{cosh}\left(\left(\frac{\beta}{2}\!-\!\tau\right)\omega\right)}{\operatorname{sinh}\left(\frac{\beta}{2}\omega\right)}\frac{\omega(W(\omega\to\infty)-W(\omega=0))}{2}
\end{equation}
and a quantity
\begin{equation}
  \widetilde{W}(\omega)=\frac{-\Im W(\omega)}{\pi}\frac{2}{\omega(W(\omega\to+\infty)-W(\omega=0))}\quad.
\end{equation}
This quantity is positive and normalized,
\begin{equation}
  \int_0^\infty\operatorname{d}\widetilde{W}=1\quad,
\end{equation}
due to the Kramers-Kronig relation
\begin{equation}
  W(\omega=0)=W(\omega\to\infty)+2\int_0^\infty\operatorname{d}\omega\frac{\Im W(\omega)}{\pi\omega}\quad.
\end{equation}
The form \eqref{eqn:bosonmaxent} together with the above properties is necessary for the applicability of the MaxEnt formalism, which is based on an interpretation of the spectral function as a probability distribution~\cite{Bryan90Maximum}. 

Casula \textit{et al.}~\cite{Casula12Dynamical} have suggested to use a factorization of the Green's function into a bosonic factor $F(\tau)$ and a part with static interaction $G_0(\tau)$, which is exact in the dynamic atomic limit approximation (DALA), 
\begin{align}
  G(\tau)=&G_0(\tau)F(\tau)\\
  =&G_0(\tau)\exp{\frac{1}{\beta}\sum_{n\neq0}\frac{U(i\omega_n)-U(0)}{\omega_n^2}(\exp{i\omega_n\tau}-1)}\quad.
\end{align}
With such a factorization, the analytic continuation of $G(\tau)$ can be separated into the analytic continuation of $G_0(\tau)$ to $A(\omega)$ and $F(\tau)$ to $B(\varepsilon)$.
The full spectral function is obtained as
\begin{equation}
  A(\omega)=\int_0^\infty\!\!\operatorname{d}\varepsilon B(\varepsilon)\frac{1+\exp{-\beta\omega}}{(1+\exp{\beta(\varepsilon-\omega)})(1-\exp{-\beta\varepsilon})}A_0(\omega-\varepsilon)\;.
\end{equation}
However, we found that this approach does \emph{not} work in the present materials simulation, which is far from the atomic limit.
It would in some cases require to do an analytic continuation of a $G_0(\tau)$ with non-causal properties, which is not possible with MaxEnt and can lead to unexpected behavior in Pad\'e.
Instead, we focused on the qualitative stability of the standard MaxEnt results with respect to the input data.
This includes robustness towards changes in statistics, error estimates, the target frequency mesh and stability over several subsequent $GW$+EDMFT steps after self-consistency is reached.

Special care must be taken to choose the real frequency range of the analytic continuation large enough that it includes all the spectral weight.
Due to the cRPA frequency-dependent bare interaction on the tier~\rom{2}, this includes very high energy plasmonic satellites, so that a range of $-150$ eV$<\omega<150$ eV was necessary.

For the $\vec{k}$-resolved spectra, the analytic continuation was done for every $\vec{k}$-point independently.